\documentclass[twocolumn, prl]{revtex4-1}
\usepackage{amssymb}

\usepackage{xcolor}
\usepackage{graphicx}
\usepackage{epstopdf}
\usepackage{color}
\usepackage{soul}
\usepackage{textcomp}
\usepackage{lineno}
\usepackage{hyperref}
\hypersetup{
	bookmarks=false,		
	unicode=false,			
	pdftoolbar=false,		
	pdfmenubar=false,		
	pdffitwindow=false,		
	pdfstartview={FitH},	
	pdftitle={null},		
	pdfauthor={null},		
	pdfsubject={null},		
	pdfcreator={null},		
	pdfnewwindow=true,		
	colorlinks=true,		
	linkcolor=blue,			
	citecolor=blue,			
	filecolor=blue,			
	urlcolor=blue			
}
\usepackage{multirow}

\begin{document}
	\title{Near-Field Scanning Microwave Microscopy in the Single Photon Regime}
	
	\author{S.~Geaney$^{1,2\dagger}$}
	\author{D.~Cox$^3$}
	\author{T.~H\"onigl-Decrinis$^1$}
	\author{R.~Shaikhaidarov$^2$}
	\author{S.~E.~Kubatkin$^4$}
	\author{T.~Lindstr\"om$^1$}
	\author{A.~V.~Danilov$^4$ and}
	\author{S.~E.~de Graaf$^{1\ddagger}$}

	\affiliation{$^1$National Physical Laboratory, Hampton Road, Teddington, TW11 0LW, UK}
	
	\affiliation{$^2$Royal Holloway, University of London, Egham, TW20 0EX, UK}
	
	\affiliation{$^3$Advanced Technology Institute, The University of Surrey, Guildford, GU2 7XH, UK}
	
	\affiliation{$^4$Department of Microtechnology and Nanoscience, Chalmers University of Technology, SE-412 96 G\"oteborg, Sweden}
		
	\affiliation{$^\dagger$\rm shaun.geaney@npl.co.uk \& $^\ddagger$sdg@npl.co.uk}
		
\begin{abstract}
The microwave properties of nano-scale structures are important in a wide variety of applications in quantum technology. Here we describe a low-power cryogenic near-field scanning microwave microscope (NSMM) which maintains nano-scale dielectric contrast down to the single microwave photon regime, up to $10^{9}$ times lower power than in typical NSMMs. We discuss the remaining challenges towards developing nano-scale NSMM for quantum coherent interaction with two-level systems as an enabling tool for the development of quantum technologies in the microwave regime.

\centering\today	
\end{abstract}
		
\maketitle

\ifx\killheadings\undefined
\section{Introduction}
\fi

Since the advent of scanning tunnelling microscopy (STM)~\cite{Binnig1982} and atomic force microscopy (AFM)~\cite{Binnig1986} a wide range of derived scanning probe microscopy (SPM) characterisation techniques have been developed, capable of nanoscale spatial mapping of a broad range of physical quantities (see e.g.~\cite{Bonnell2012} for a review). The rapid development of nanotechnology, materials and surface science underpinned by these techniques drives the demand for ever more versatile and non-invasive nanoscale analysis tools. In particular, for the rapidly growing field of quantum device technologies there is a need to develop supporting SPM techniques operating in the same regime as these devices themselves \textit{i.e.} in the quantum coherent regime. However, the number of nanoscale characterisation tools capable of quantum coherent interaction with samples are so far very limited~\cite{Bonnell2012, Meyer2013, Shanks2013, Tisler2013, Balasubramanian2008}. In particular, at microwave frequencies where photon energies are orders of magnitude smaller than for optical wavelengths this poses a tremendous challenge due to the lack of single photon detectors and strict requirement for millikelvin temperatures. However, tools operating in this regime are urgently needed to drive further developments in solid state quantum technologies. 

Near-field scanning microwave microscopy (NSMM) combines microwave characterisation with either STM~\cite{Knoll1997} or AFM~\cite{Wu2018} using either a broadband~\cite{Vlahacos1998} or resonant~\cite{Gregory2014} probe. In the near-field mode the spatial resolution is limited by the size of the SPM tip which can be many orders of magnitude below the diffraction limit. Various implementations of NSMM has been used extensively in the classical regime to non-invasively obtain surface and subsurface information on semiconductor devices~\cite{Kundhikanjana2011}, defects in 2D materials~\cite{Gregory2014}, biological samples~\cite{Tuca2016} and for investigating high-$T_c$ superconductivity~\cite{Lann1999}, to name a few applications (for an overview see e.g.~\cite{Anlage2007}).

\begin{figure}[b]
	\centering
	\includegraphics[width=.45\textwidth]{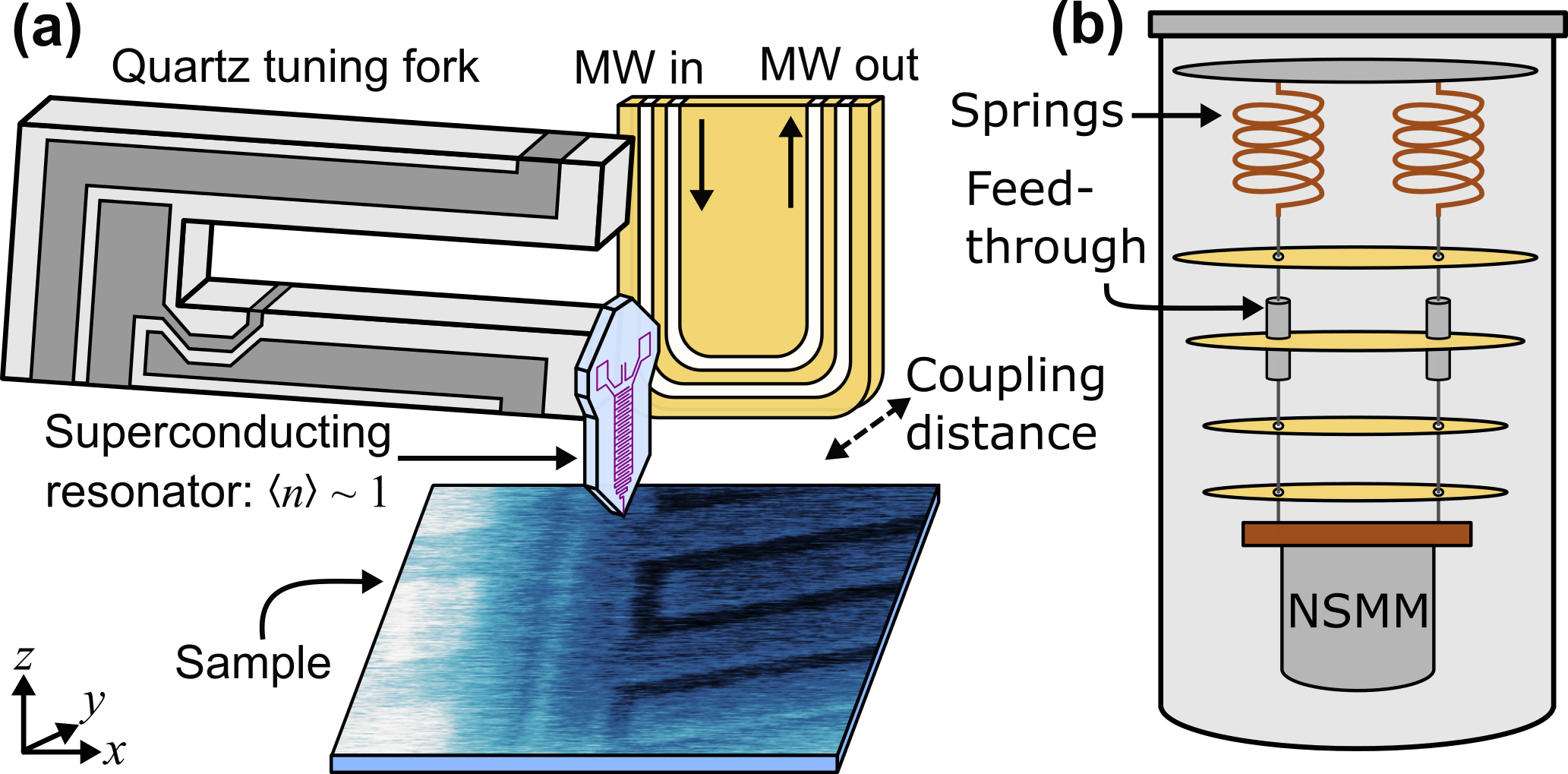}
	\caption{\textbf{(a)} Diagram of the set-up of the NSMM microscope. A superconducting fractal resonator, with average photon occupancy $\langle n \rangle \sim 1$, is adhered to a quartz tuning fork. A movable CPW is used to inductively couple to the resonator for excitation and readout. \textbf{(b)} Illustration of the NSMM suspended by springs and kevlar wire inside a dilution refrigerator. Feed-throughs at the 800~mK plate are designed to reduce the amount of thermal photons from hotter stages reaching the NSMM.}
	\label{fig:setupNSMM}
\end{figure}

Here we present the first NSMM that operates in the single microwave photon regime and at 30 mK. We show that our NSMM is capable of obtaining nano-scale dielectric information in this regime of ultra-low power.

This is an important step towards developing the future tool kit for characterisation of solid state quantum circuits and truly non-invasive nanoscale microwave interrogation of quantum materials and devices. Operating the NSMM in the quantum limit enables a range of applications when the microwave signals of the NSMM can coherently couple quantum two-level systems (TLS). This includes nanoscale quantum materials characterisation in the microwave domain, understanding the nature of individual microscopic two-level system defects in quantum devices~\cite{Muller2017, deGraaf2015} (those responsible for qubit parameter fluctuations~\cite{Burnett2019, Schlor2019, Klimov2018}), interrogation of engineered TLS (such as quantum dots and superconducting qubits), individually or in large-scale quantum circuits, probing valley physics in silicon, and probing the physics in quantum metamaterials~\cite{Zagoskin2016, Burkard2019, Jung2014}. We discuss the limitations of our NSMM and the current limitations towards enabling it for these applications. We not that this technique could also be used to study local field distributions~\cite{Feenstra1998}, quasiparticle dynamics and loss~\cite{Grunhaupt2018} and noise~\cite{Burnett2014, deGraaf2017, deGraaf2018} mechanisms in e.g. superconductors and dielectrics in the ultra-low power regime.

For coherent nanoscale measurements with NSMM there are four main requirements to satisfy: (i) The microscope must operate at low temperatures to ensure that the thermal energy is much less than the probe frequency (and any TLS energy level splitting  $E_{\text{TLS}}$, of interest), $k_B T \ll E_{\text{TLS}} \sim \hbar\omega_r$, where $\omega_r$ is the NSMM probe angular resonance frequency. For $\omega_r/2\pi = 6$~GHz this requires $T \ll 300$ mK. (ii) The NSMM needs to operate at low microwave powers to be able to coherently couple to TLS and without saturating them~\cite{Stoutimore2012}. The critical photon number for saturation in the dispersive regime is given by $\eta_c = (\omega_r - E_{\text{TLS}}/\hbar)^2/ 4 g^2$, where $g$ is the coupling strength~\cite{Boissonneault2008}, which implies that the average number of photons in the resonator $\langle n \rangle \ll \eta_c$, must be close to one \textit{i.e.} the NSMM should operate in the near single photon regime. (iii) The resonator loss rate $Q_i^{-1}$ should be smaller than the coupling strength $g$, requiring a high-$Q$ resonator. (iv) Nanometer scale distance control between the tip and the sample surface is needed for a well defined coupling between the probe and a TLS, prompting the integration with AFM in a system that is well isolated from vibrations. Here we demonstrate the first NSMM making significant advances towards reaching all these stringent requirements while still maintaining the capability of nanoscale dielectric imaging.

\ifx\killheadings\undefined
\section{Experiment}
\fi
To achieve precise distance control between the tip and the sample and to enable scanning the tip across the sample surface we use tuning-fork based AFM. This technique is compatible with a cryogenic environment due to its low dissipation electrical readout~\cite{Senzier2007}. The use of quartz tuning forks for AFM is a well established technique~\cite{Karrai1995, Wu2018} which can be used to perform non-contact SPM on both conducting and dielectric surfaces. The tuning fork response is read out electrically using a capacitance compensating circuit~\cite{Grober2000} and a low noise pre-amplifier. A phase-locked loop (PLL) in our SPM controller measures the tuning fork resonance frequency shift and provides the AFM feedback. The bare tuning fork has a resonance frequency of $f_{r,\text{TF}} = 32.7$~kHz.

The experimental set-up of the NSMM is shown in FIG.~\ref{fig:setupNSMM}(a). To integrate microwave interrogation with the AFM we use a thin-film Nb microwave resonator ($f_r = \omega_r/2\pi \approx 6$~GHz) patterned onto a silicon substrate and micromachined into a small resonator that is adhered to a single prong of a tuning fork~\cite{deGraaf2015}. The resonator is designed to be small and compact so that it can fit onto the end of a tuning fork whilst maintaining the properties of a distributed resonator~\cite{deGraaf2013}. The resonator is terminated with the AFM tip placed at a microwave voltage anti-node. Typically we find $f_{r,\text{TF}} \sim$ 29 -- 30 kHz with a $Q$-factor of $\sim 10^4$ at 30~mK due to the added mass of the microwave resonator on one of the prongs. To make a well-defined tip we use xenon focused ion beam (FIB) etching from the backside of the resonator (to prevent damage to the resonator patterned on the front side) to mill the tip to the desired size. Milling with the Xe-FIB from the backside (non-metalised side) of the probe has no observed impact on the quality factor of the microwave resonator. The FIB step is required for high resolution and high sensitivity NSMM imaging, ensuring the NSMM tip is well defined and that the AFM and metallic NSMM tips are one and the same. The resonator is excited and read-out through a co-planar waveguide (CPW) made from a printed circuit board (PCB) which inductively couples to the resonator. The coupling strength (distance between resonator and coupling PCB) can be tuned with the use of an Attocube ANPz30 piezo-stepper that is placed behind the CPW, allowing for optimisation of the coupling of the resonator. Coarse positioning of the probe above the sample is achieved with Attocube ANC150 piezo-steppers whereas fine-movements and scanning is done with a set of in-house built piezo-tube scanners with a scan range of $18$ \textmu m at 30 mK.

The whole set-up is housed within custom casing and is suspended in a BlueFors LD-400 dilution refrigerator (See FIG.~\ref{fig:setupNSMM}(b)) from three copper-beryllium (CuBe) springs and kevlar thread that feeds through from the 50~K plate to the mixing chamber plate. This aims to minimise the effects of external mechanical vibration by acting as a mechanical low-pass filter, in particular to reduce the vibrations caused by the pulse tube operating at 1.4~Hz. The total suspended mass is 5~kg and with a combined spring constant of $k = 295$~N/m this results in a resonant frequency of the suspension of 1.2~Hz. This is crucial as vibrations will significantly impact the distance control and the performance of the coherent NSMM. At the 800~mK (still) plate we designed a feed-through for each kevlar thread to reduce the amount of thermal photons reaching the NSMM, while maintaining the mechanical properties of the suspension. The suspension feed-through are thin hollow tubes with inner walls painted in stycast with a suspended baffle over the top that block a direct line-of-sight from the top to the bottom of the fridge.

\begin{figure*}[t]
	\centering
	\includegraphics[width=.95\textwidth]{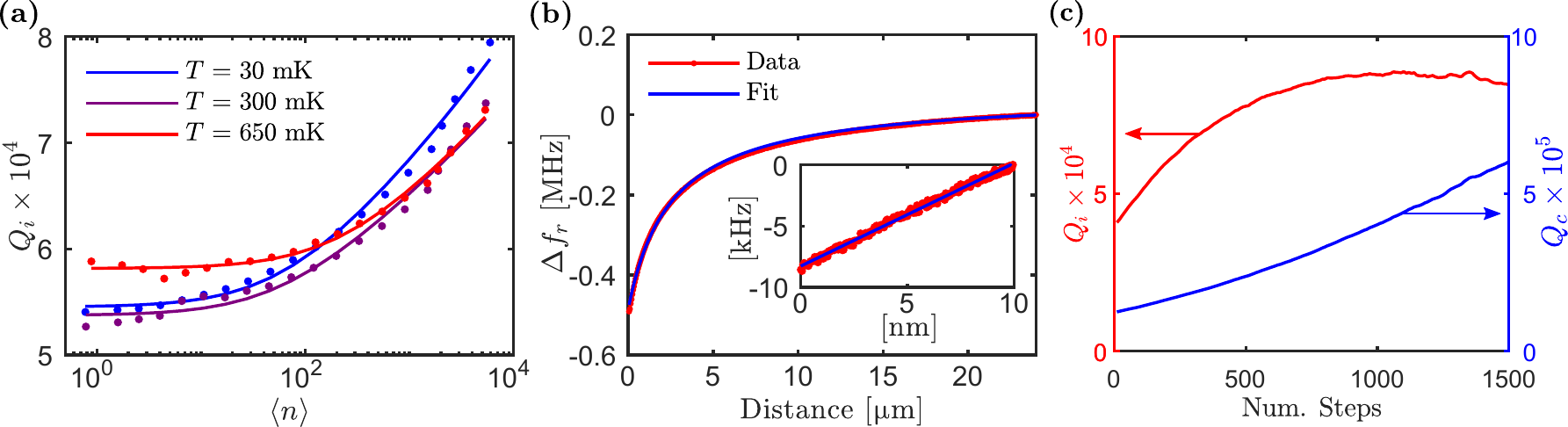}
	\caption{\textbf{(a)}. The intrinsic quality factor of the resonator probe as a function of average photon number $\langle n \rangle$, for three different temperatures. The data ($\bullet$) is fitted (\textbf{---}) to $Q_i(\langle n\rangle)$ (see text). \textbf{(b)} The frequency shift of the resonator probe as a function of tip-to-sample distance. Inset: The frequency shift at small distances with a linear approximation. The gradient of 0.84~kHz/nm from this data is what we use to convert from frequency noise to tip-to-sample displacement noise. \textbf{(c)}. The intrinsic (left axis) and coupling (right axis) quality factors as a function of the number of steps made by the coupling piezo-stepper (coupling distance).}
	\label{fig:QvsP_frvsD_QiQcvsStep}
\end{figure*}

To measure the real-time frequency shift of the superconducting resonator with high-sensitivity we employ the Pound-Drever-Hall (PDH) technique. This method is commonly used in optics for laser frequency stabilisation~\cite{Black2001}. Here this method gives us the ability to accurately monitor the microwave resonance frequency while scanning the tip over the sample surface. Another advantage of the PDH loop is that it is immune to variations in the electrical length caused by thermal drift, moving parts and other noise processes, making it an ideal technique for NSMM, as opposed to an interferometric technique such as homodyne detection. In brief, the PDH method uses a carrier tone near the resonant frequency $f_c$, that is phase modulated with a frequency $f_m$. The side-bands at $f_c \pm f_m$ are far detuned from resonance and therefore do not interact with the resonator, while $f_c$ acquires a phase shift proportional to the detuning $\delta f = f_c - f_r$ from the instantaneous resonance frequency $f_r$. Passing this spectrum through the resonance and to a non-linear detector produces a beating signal at $f_m$ with an amplitude proportional to the detuning $\delta f$. A PID controller aims to null this beating signal, thus tracking the resonance frequency. A detailed discussion on the PDH technique can be found in Ref.~\cite{Lindstrom2011}.

The use of the PDH technique for NSMM was originally outlined in~\cite{deGraaf2013}. Here we use the same general set-up but with significantly improved cryogenic microwave circuitry to facilitate measurements in the single photon regime. A lock-in amplifier provides the phase modulation reference signal at $f_m$ which is combined with the carrier in a phase modulator to produce the phase modulated spectrum sent to the cryostat. The amplified and filtered output signal from the cryostat is measured with a diode detector and the detector output is lock-in demodulated with $f_m$ as a reference. The lock-in output is fed to a PID controller which in turn controls the carrier frequency $f_c$ through the frequency modulation input of the microwave generator.

\ifx\killheadings\undefined
\section{Results \& Discussion}
\fi
One of the requirements for coherent NSMM is to operate at low power to reach the single photon regime~\cite{deGraaf2015}. To evaluate the average photon number we measure the intrinsic quality factor $Q_i$, of the resonator probe as a function of applied microwave power using a Vector Network Analyser (VNA). This data is shown in FIG.~\ref{fig:QvsP_frvsD_QiQcvsStep}(a) for three different temperatures. For a resonator with impedance close to 50~$\Omega$ we get the corresponding average photon number $\langle n \rangle$, for a given input power $P_{\text{in}}$, using the equation
\begin{equation}
\langle n \rangle = \frac{\langle E_{\text{int}} \rangle}{\hbar \omega_r} = \frac{2}{\pi}\frac{Q^2}{Q_c}\frac{P_{\text{in}}}{\hbar\omega_r^2},
\label{eq:photonNumber}
\end{equation}
where $Q_c$ is the coupling quality factor and $Q$ is the total quality factor. The data for $Q_i(\langle n \rangle)$ shown in FIG.~\ref{fig:QvsP_frvsD_QiQcvsStep}(a) has been fitted to a power law dependence adapted from the standard tunnelling model that states $Q_i^{-1} = F\tan(\delta_i)/(1+ \langle n \rangle / n_c)^\alpha + Q_{i,0}^{-1}$ ~\cite{Phillips1987} where $Q_{i,0}^{-1}$ accounts for losses that are independent of power, $F$ is the filling factor of the TLS hosting medium in the resonator, $n_c$ is a critical number of photons for saturation, $\tan(\delta_i)$ is the loss tangent; the fit returns $\alpha = 0.07$ and $F\tan(\delta_i) = 1.8 \times 10^{-5}$ at 30~mK, to our knowledge the highest NSMM $Q$ reported. FIG.~\ref{fig:QvsP_frvsD_QiQcvsStep}(a) shows that at lower photon numbers $Q_i$ saturates in the single photon regime at a power-independent value as expected for TLS related dissipation.

Next, we investigate the resonator behaviour as it is moved in close proximity to the sample surface. In FIG.~\ref{fig:QvsP_frvsD_QiQcvsStep}(b) we show measured data of the resonant frequency shift of the NSMM probe as a function of tip-to-sample distance. This shift is due to the changing capacitance $\Delta C$ between the metallic probe tip and the sample. We fit the data using the equation $\Delta\omega_r = 1/\sqrt{L(C + \Delta C)} - 1/\sqrt{LC}$, where $L$ and $C$ are the inductance and capacitance of the resonator respectively. We calculate $\Delta C$ by assuming the tip is a metallic sphere of radius $R$ at a distance $z$ above an infinite conducting plane~\cite{Hudlet1998}
\begin{equation}
\Delta C = 2\pi\varepsilon_0 R \ln\left(1 + \frac{R}{z + \Delta z}\right).
\end{equation}
We add an offset distance $\Delta z$, to account for the fact that the data in FIG.~\ref{fig:QvsP_frvsD_QiQcvsStep}(b) starts at an unknown distance away from the surface. From this analysis we find the microwave tip radius $R = 2$~\textmu m. To convert between frequency and distance in NSMM scans, we measure the shift in the resonant frequency as a function of tip-to-sample distance close to the sample surface (inset of FIG.~\ref{fig:QvsP_frvsD_QiQcvsStep}(b)). We find a linearised frequency-to-distance conversion coefficient of 0.84~kHz/nm.

We then measure the effect of adjusting the coupling distance between the resonator and the CPW. We do this find the optimal coupling to the resonator, which may vary between probe assemblies. FIG.~\ref{fig:QvsP_frvsD_QiQcvsStep}(c) shows the change in $Q_i$ (left axis) and $Q_c$ (right axis) as the distance between the CPW and the resonator probe is changed. As expected, $Q_c$ increases with increased coupling distance. The response in $Q_i$ follows from the change in $\langle n \rangle$ induced by the change in $Q_c$: combining Eq.~\ref{eq:photonNumber} with $Q_i(\langle n\rangle)$ we have in the limit $Q_i \gg Q_c$ that $Q_i \propto Q_c$ and in the limit $Q_i \ll Q_c$ that $Q_i \propto Q_c^{-\alpha/(1-2\alpha)} = Q_c^{-0.08}$ for $\alpha = 0.07$, in good agreement with the measured change in $Q_i$.

\begin{figure}[b]
	\centering
	\includegraphics[width=.46\textwidth]{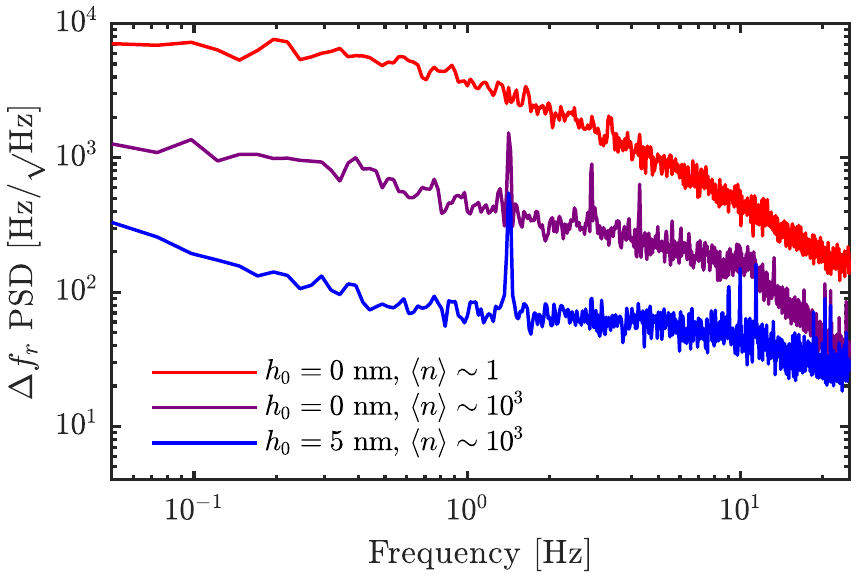}
	\caption{PSD of the microwave frequency shift $\Delta f_r$, from the superconducting resonator. The peak at 1.4~Hz is the pulse tube. Red line: When the tip is in contact at low power ($\langle n \rangle \sim 1$). Purple line: When the tip is in contact at high power. Blue line: When the tip is retracted 5~nm from the surface at high power. The kink in the data from $\sim 10^1$ Hz is the roll-off of the PID bandwidth in the PDH loop.} 
	\label{fig:PSDplots}
\end{figure}

Next, to determine the mechanical stability of the NSMM we measure the power spectral density (PSD) of the fluctuation in centre frequency of the resonator $\Delta f_r$ for three different parameter combinations of tip-to-sample distance and average photon number, specified in FIG.~\ref{fig:PSDplots}. There are peaks at 1.4~Hz and harmonics thereof due to the dilution refrigerator pulse tube on two of the PSD traces. In the high power regime ($\langle n \rangle \sim 10^3$) the frequency fluctuations of the microwave resonator are limited by the mechanical noise of the system, which translates to frequency noise through the fluctuations in tip-sample capacitance. When the NSMM probe is in contact with the sample, the peak amplitude of 1.5~kHz/$\sqrt{\text{Hz}}$ at 1.4~Hz corresponds to 1.8~nm. We note that despite the relative simplicity of the suspension inside our dry dilution refrigerator these noise levels are very much comparable to other similar state of the art scanning probe microscopes~\cite{Pelliccione2013, Shperber2019}. Lifting the tip by 5~nm reduces the sensitivity of the resonator to mechanical noise, indicating that the mechanical noise limits the frequency read-out accuracy of the microwave resonator at high powers. However, in the single photon limit the noise level is much higher and the peaks due to the pulse tube are washed out. Here the dominating noise process at the time-scales shown in FIG.~\ref{fig:PSDplots} is the white noise of the measurement set-up. The $1/f$ intrinsic noise level of the resonator due to TLS defects~\cite{Burnett2014} was independently found to be $\sim 380~\text{Hz}/\sqrt{\text{Hz}}$ (at $f = 1$~Hz) for $\langle n \rangle \sim 1$ and is comparable to the mechanical noise.

\begin{figure*}[t]
	\centering
	\includegraphics[width=.95\textwidth]{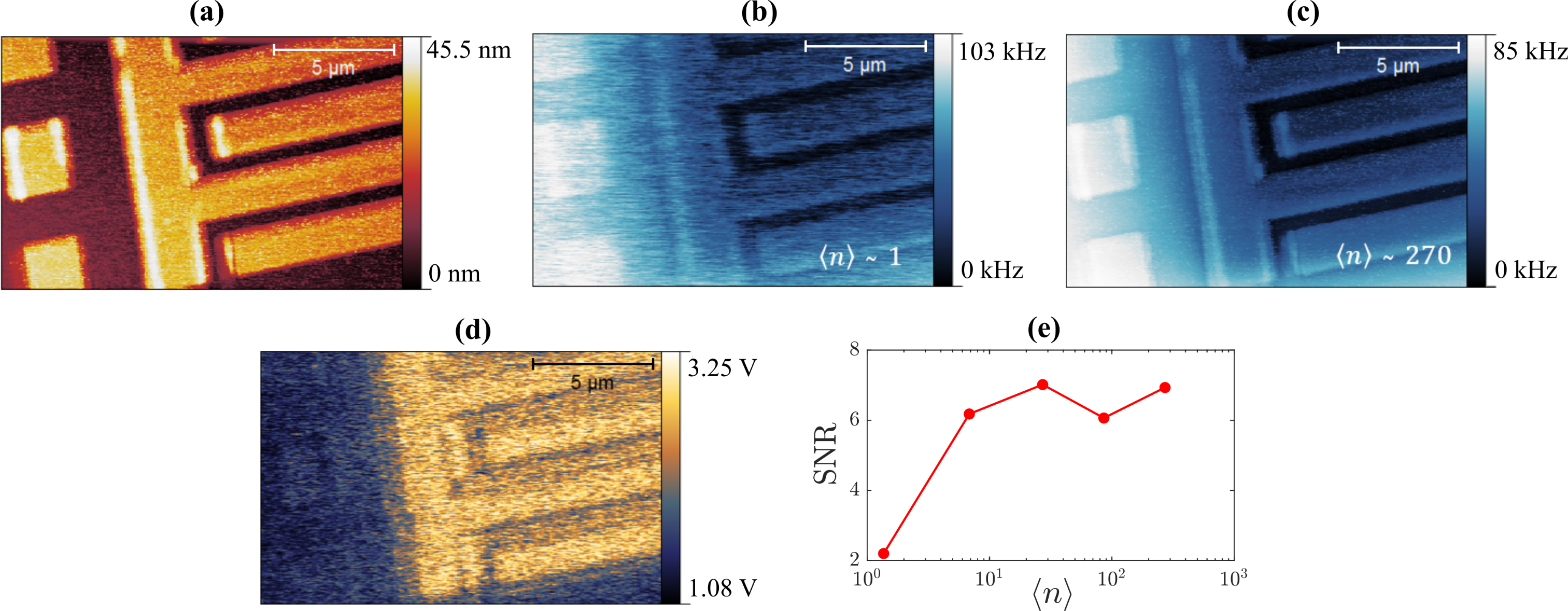}
	\caption{Scans of an interdigitated capacitor with adjacent metallic pads taken at 30~mK. \textbf{(a)} AFM Topography scan. \textbf{(b)} Single photon regime microwave scan ($\langle n \rangle \sim 1$) showing the frequency shift of the microwave resonator. The images were acquired using a scan speed of 0.67~\textmu m/s. \textbf{(c)} Microwave scan at high power ($\langle n \rangle \sim 270$). \textbf{(d)} The PDH error signal demodulated at the tuning fork frequency (30~kHz), proportional to $df_r/dz$ ($\langle n \rangle \sim 270$). \textbf{(e)} The signal-to-noise ratio (SNR) obtained from scans as a function of the average photon number $\langle n \rangle$.}
	\label{fig:scans}
\end{figure*}

Finally, we demonstrate scanning with nanoscale resolution in the single photon regime, shown in FIG.~\ref{fig:scans}. The scans are of the same area of a sample consisting of Al patterned on a Si substrate. The scan shows three metallic squares ($2 \times 2$~\textmu m$^2$) placed adjacent to two larger structures that form an interdigitated capacitor. Each metal finger of the interdigitated capacitor has a width and separation of 1~\textmu m, although in FIG.~\ref{fig:scans} these distances appear different due to the shape of the tip.

FIG.~\ref{fig:scans}(a), shows the AFM topography, and the microwave resonator response, taken in the single photon regime, is shown in FIG.~\ref{fig:scans}(b). Remarkably, even at these ultra-low power levels, up to $10^{9}$ times lower than in conventional NSMMs (reported power levels in the literature are down to about $-20$~dBm~\cite{Wu2018}), we can resolve a clear contrast in the NSMM image. A similar scan taken at higher powers ($\langle n \rangle \sim 270$) is shown in FIG.~\ref{fig:scans}(c). As expected, the scan taken in the single photon regime is noisier than the equivalent high power scan, see signal-to-noise ratio (SNR) plot in FIG.~\ref{fig:scans}(e).

As the tip-to-sample distance is kept constant by the AFM feedback, the contrast shown in the microwave scans is therefore mainly due to changes in capacitance between the tip and the sample. The smaller metallic squares in the scan are brighter than the larger metallic structures since a smaller structure has a weaker capacitive coupling to ground compared to larger ones.

This is further supported by FIG.~\ref{fig:scans}(d) that shows the response of the resonator at $\langle n \rangle \sim 270$, demodulated at the tuning fork frequency of 30~kHz. The contrast originates from the change in microwave resonance frequency as the tip oscillates at the tuning-fork frequency in close proximity to the sample surface. The same scan was done at $\langle n \rangle \sim 1$ but found to have a SNR less than 1. The demodulated signal is the PDH loop `error' signal which is not tracked by the PID (which only has a bandwidth up to $\sim$10~kHz). For variations smaller than the resonance linewidth the `error' signal becomes directly proportional to the linearised phase response around $f_{r}$ and thus the demodulated signal is proportional to $df_r/dz$.

FIG.~\ref{fig:scans}(d) highlights the contrast formation mechanism in FIG.~\ref{fig:scans}(b) and FIG.~\ref{fig:scans}(c) originating from the capacitive network formed between tip, sample features and ground plane which is on the backside of the silicon substrate. The reduced contrast for the smaller metallic squares implies that the size of the tip and its own capacitance to ground dominates the total capacitance, whereas the larger metallic structures have a much larger self-capacitance, resulting in the response being dominated by the smaller time-dependent tip-sample capacitance. This cross-over occuring at length-scales of $\sim$~1~\textmu m is in good agreement with the size of the near-field tip estimated from the approach curve in FIG.~\ref{fig:QvsP_frvsD_QiQcvsStep}(b).

Microwave scans like the ones shown in FIG.~\ref{fig:scans} were performed at several different average photon numbers. From these we calculated a signal-to-noise ratio (SNR) as a function of the average photon number, shown in FIG.~\ref{fig:scans}(d). The difference between the average response on a metallic area and on an area of dielectric substrate is used to evaluate the signal. We divide this signal by the noise, estimated as the root mean square variation over the same areas. As expected, the SNR is lower for scans close to the single photon limit than for higher power scans. This is in agreement with the previous conclusion that, except for the scans taken at very low average photon numbers, the noise is dominated by mechanical noise which is independent of the applied microwave power.

In the single photon regime we would expect to be capable of coupling to coherent quantum devices in the microwave domain~\cite{deGraaf2015}. For example, even on the surface of naturally oxidised Al one would expect to observe $\sim$~1 TLS defect per \textmu m$^3$ within a 100~kHz bandwidth around the resonance frequency~\cite{Burnett2016, Klimov2018}. Currently our NSMM requires some further improvements to be able to coherently detect TLS. As stated before, mechanical stability is currently not a limiting factor in the single photon regime. The main limiting factor is instead revealed in the temperature dependence of the single photon $Q$-factor (FIG.~\ref{fig:QvsP_frvsD_QiQcvsStep}(a)). (This was separately confirmed through the temperature dependence of the resonance frequency (data not shown)). The change in $Q_i(T)$ is much smaller than expected indicating that the TLS defect bath and the resonator are not entirely thermalised. From the thermal saturation of two-level system defects in the resonator we expect $Q_i(T) \propto \tanh(\hbar\omega/k_B T)$~\cite{Burnett2017}. For our measured temperature range this translates into the ratio $Q_i(T = 30~\text{mK})/Q_i(T = 650~\text{mK}) \sim 4.5$. However, we observe a ratio of $\sim 1.5$, indicating that the TLS bath is much warmer than the mixing chamber temperature. This was confirmed in a well-shielded separate cool-down of the same probe, which showed the expected temperature dependence. Additional engineering of the NSMM enclosure, suspension feed-throughs, and thermal anchoring is needed to ensure increased thermalisation.

\ifx\killheadings\undefined
\section{Conclusions}
\fi
We have designed and evaluated a resonant near-field microwave microscope operating at 6~GHz and 30~mK. The microscope for the first time demonstrates nano-scale dielectric contrast in the single microwave photon regime. Our results shows promise for the development of microwave SPM instrumentation capable of coherently interacting with quantum systems. We discussed the main engineering challenges to our current set-up in achieving this significant milestone. A milestone that would enable an entirely new tool to characterise and drive the development of quantum devices and technologies in the microwave domain.

\ifx\killheadings\undefined
\section{Acknowledgements}
\fi
We thank A. Ya. Tzalenchuk for careful reading of the manuscript as well as P. J. Meeson and J. Burnett for their helpful discussions during this work. This work was supported by the UK Department of Business, Energy and Industrial Strategy (BEIS). AD acknowledges support from the Swedish Research Council (VR), grant \# 2016-048287.


\renewcommand\thefigure{\thesection S\arabic{figure}}
\setcounter{figure}{0} 
\renewcommand\theequation{S\arabic{equation}}
\setcounter{equation}{0} 

\end{document}